\documentclass[a4paper,11pt]{article}
\usepackage{pos}

\title{Neutrino Masses and Hubble Tension via a Majoron in MFV}

\author[a]{Fernando Arias-Aragón}
\author[b,c]{Enrique Fernández-Martínez}
\author*[b,c]{Manuel González-López}
\author[b,c]{Luca Merlo}

\affiliation[a]{Laboratoire de Physique Subatomique et de Cosmologie, Université Grenoble-Alpes\\
CNRS/IN2P3, Grenoble INP, 38000, Grenoble, France}

\affiliation[b]{Instituto de Física Teórica UAM-CSIC,\\
 Calle Nicolás Cabrera 13-15, Cantoblanco, 28049 Madrid,  Spain}

\affiliation[c]{Departamento de Física Teórica, Universidad Autónoma de Madrid\\
Cantoblanco, 28049, Madrid, Spain}

\emailAdd{fernando.arias-aragon@lpsc.in2p3.fr}
\emailAdd{enrique.fernandez-martinez@uam.es}
\emailAdd{manuel.gonzalezl@uam.es}
\emailAdd{luca.merlo@uam.es}

\abstract{The recent tension between local and early measurements of the Hubble constant can be explained in a particle physics context. A mechanism is presented where this tension is alleviated due to the presence of a Majoron, arising from the spontaneous breaking of Lepton Number. The lightness of the active neutrinos is consistently explained.  Moreover, this mechanism is shown to be embeddable in the Minimal (Lepton) Flavour Violating context, providing a correct description of fermion masses and mixings, and protecting the flavour sector from large deviations from the Standard Model predictions. A QCD axion is also present to solve the Strong CP problem. The Lepton Number and the Peccei-Quinn symmetries naturally arise in the Minimal (Lepton) Flavour Violating setup and their spontaneous breaking is due to the presence of two extra scalar singlets. The Majoron phenomenology is also studied in detail. Decays of the heavy neutrinos and the invisible Higgs decay provide the strongest constraints in the model parameter space.}

\FullConference{%
  *** The European Physical Society Conference on High Energy Physics (EPS-HEP2021), ***\\
  *** 26-30 July 2021 ***\\
  *** Online conference, jointly organized by Universität Hamburg and the research center DESY ***
}

\begin{document}
\maketitle

\section{Introduction}
Currently,  local observations and early Universe measurements disagree in the determination of the expansion rate of the Universe.  The so-called Hubble tension stands at the level of 4-6$\sigma$ \cite{verde,wong}. The solution to this problem may rely on systematics in the observations or lead to a new cosmological model, but it can also be tackled from the point of view of particle physics~\cite{nosotros}.

Ref.~\cite{miguel} suggested that the existence of a Majoron ($\omega$) coupled to light neutrinos may significantly alleviate the Hubble tension,  reducing it to 2.5$\sigma$.  In order to do so, the Majoron mass\footnote[1]{The Majoron mass, possibly arising from gravitational effects, is an open research topic. We will remain agnostic to its origin, assuming it to lie in the range of interest.} and coupling must lie in the ranges
\begin{equation}
m_\omega \in[0.1,1]\,\mathrm{eV},\enspace\lambda_{\omega\nu\nu}\in[5\times 10^{-14},10^{-12}]\,.
\label{eq:ranges}
\end{equation}
The Majoron is the Goldstone boson associated to the spontaneous breaking of lepton number (LN), which is only an accidental symmetry within the SM.  This new particle could naturally appear in a type-I seesaw mechanism, and, provided it carries lepton number,  be responsible for the generation of Majorana masses via a Yukawa coupling to neutrinos. 

\section{The Majoron mechanism}
We extend the SM particle content by introducing three right-handed neutrinos (RHN) and a new scalar, $\chi$, which are singlets under the SM gauge group.  All these new particles are charged under LN, with the concrete assignments ($-L_N$ and $L_\chi$, respectively), being a priori free.  In this setup, the most general neutrino Lagrangian can be written as
\begin{equation}
-\mathcal{L}_\nu=\left(\frac{\chi}{\Lambda_\chi}\right)^{\frac{1+L_N}{L_\chi}}\bar{\ell}_L\tilde{H}\mathcal{Y}_\nu N_R + \frac{1}{2}\left(\frac{\chi}{\Lambda_\chi}\right)^{\frac{2L_N-L_\chi}{L_\chi}} \chi \bar{N}^c_R\mathcal{Y}_NN_R+\mathrm{h.c.}
\end{equation}
where $H$ is the SM Higgs doublet, $\tilde{H}=i\sigma_2H$, $\mathcal{Y}_\nu$ and $\mathcal{Y}_N$ are Yukawa matrices, and $\Lambda_\chi$ is the cut-off scale associated to the new scalar.  Once $\chi$ and the Higgs take vevs ($v$ and $v_\chi$, respectively),  light and heavy neutrino masses, as well as a Majoron-neutrino coupling, are generated:
\begin{equation}
m_\nu=\frac{\varepsilon_\chi^{\frac{2+L_\chi}{L_\chi}} v^2}{\sqrt{2} v_\chi}\mathcal{Y}_\nu\mathcal{Y}_N^{-1}\mathcal{Y}_\nu^T,\enspace
M_N\simeq\varepsilon_\chi^{\frac{2L_N-L_\chi}{L_\chi}} \frac{v_\chi}{\sqrt{2}}\mathcal{Y}_N,\enspace
\lambda_{\omega\nu\nu}=2\dfrac{m_\nu}{L_\chi v_\chi}\,,
\end{equation}
where $\varepsilon_\chi=\frac{v_\chi}{\sqrt{2}\Lambda_\chi}$.  

Clearly, the features of the model depend crucially on the LN assignments. Although $L_N$ and $L_\chi$ are treated as free parameters,  some conditions constrain them. For instance, the combinations $(1+L_N)/L_\chi$ and $(2L_N-L_\chi)/L_\chi$ must be natural numbers in order for the Lagrangian to be local.  

We consider two main scenarios, both of them non-renormalizable. The LN charges in both of them are shown in Tab.~\ref{tab:param}, as well as the ranges in which the relevant parameters would lie ($M_N$ and $\theta$ stand for the heavy neutrinos mass and mixing,  respectively).  These are obtained by imposing the light neutrino masses to agree with observations ($m_\nu\simeq\sqrt{\Delta m^2_\mathrm{atm}}$) and the Majoron coupling to neutrinos to agree with Eq.~\ref{eq:ranges}.  In order to explore whether these features can be achived with no need to fine tune the Yukawas, we assume that the entries of the matrices $\mathcal{Y}_\nu$ and $\mathcal{Y}_N$ are $\mathcal{O}(1)$. 
\begin{table}[h]
\hspace{-0.8cm}
\begin{tabular}{|c|c|c|c|c|c|c|c|}
\hline
&&&&&&&\\[-4mm]
		& $L_N$ & $L_\chi$ & $v_\chi$ (TeV) & $\varepsilon_\chi\, (10^{-6})$ & $M_N$ (GeV) & $\Lambda_\chi$ (TeV)&$\theta^2$ \\[1mm] \hline
&&&&&&&\\[-4mm]
CASE NR1 	& $1$ & $1$ & $[0.1,2]$& $[49,140]$ & $[0.003,0.2]$ & $[1.4-11]\times10^3$&$(0.25-14)\times10^{-9}$\\[1mm] \hline
&&&&&&&\\[-4mm]
CASE NR2 	& $1$ & $2$ & $[0.05,1]$& $[0.24,1.1]$ & $[35.4,707]$ & $[1.4-6.5]\times10^5$&$(0.7-14)\times 10^{-13}$\\[1mm]
\hline
\end{tabular}
\caption{\em Parameter ranges in the two phenomenologically interesting scenarios.}
\label{tab:param}
\end{table}

We find that both cases NR1 and NR2 yield reasonable ranges for the relevant parameters of the model.  We do not consider the renormalizable case ($L_N=-1$, $L_\chi=-2$), as it is not possible to simultaneously reproduce light neutrino masses and  improve the Hubble tension without a fine-tuning of the Yukawas. The case $L_N=L_\chi=-1$ is also disregarded, as it would require $\varepsilon_\chi\gg1$. 
\section{Phenomenological signatures}
\begin{enumerate}
  \item[$\blacksquare$] \textbf{Majoron couplings. }Majorons could be emitted in neutrinoless double-beta decay processes due to their tree level coupling to neutrinos.  Current bounds on the half-life of these decays are not stringent enough, so the range in Eq.~\ref{eq:ranges} is perfectly allowed. 
  
At loop level, Majorons exhibit couplings to photons and electrons, which are respectively constrained by searches for very light pseudoscalars and astrophysical measurements based on Red Giants.  Our mechanism predicts Majoron couplings to electrons and photons which are safe from current bounds by many orders of magnitude, due to the loop suppression.
\item[$\blacksquare$]\textbf{Heavy neutrinos.} As seen in Tab.~\ref{tab:param},  both cases NR1 and NR2 predict the appearance of RHNs with testable masses.  Scenario NR1 shows the better prospects, as neutrinos in the MeV range could be detected at beam dump experiments. This case is however disfavored by cosmological observations, as the RHNs would be quite long-lived, possibly spoiling Big Bang Nucleosynthesis predictions.  This would not be a problem in case NR2, where the neutrinos are heavier and, thus, short-lived. In this case, such particles could be probed at collider facilities, although their mixing with light neutrinos would be pretty small, making their detection challenging from an experimental perspective.
\item[$\blacksquare$]\textbf{Scalar sector.} Once the new scalar takes a vev, both its angular ($\omega$, the Majoron) and radial ($\sigma$) parts appear in the spectrum. The latter will mix with the SM Higgs (due to the quartic coupling in the potential), also inducing it to invisibly decay to two Majorons.  LHC measurements constrain the Higgs mixing and invisible decay. Both limits can be translated into the parameter space $(M_\sigma,g$), where  $g$ is the strength of the quartic coupling.  We find that, for values of $v_\chi$ consistent with cases NR1 and NR2, it is possible to respect current bounds without fine-tuning the quartic coupling or requiring $\sigma$ to be too heavy.
\end{enumerate}
\begin{figure}[h]
\centering
\includegraphics[height=5.5cm,keepaspectratio]{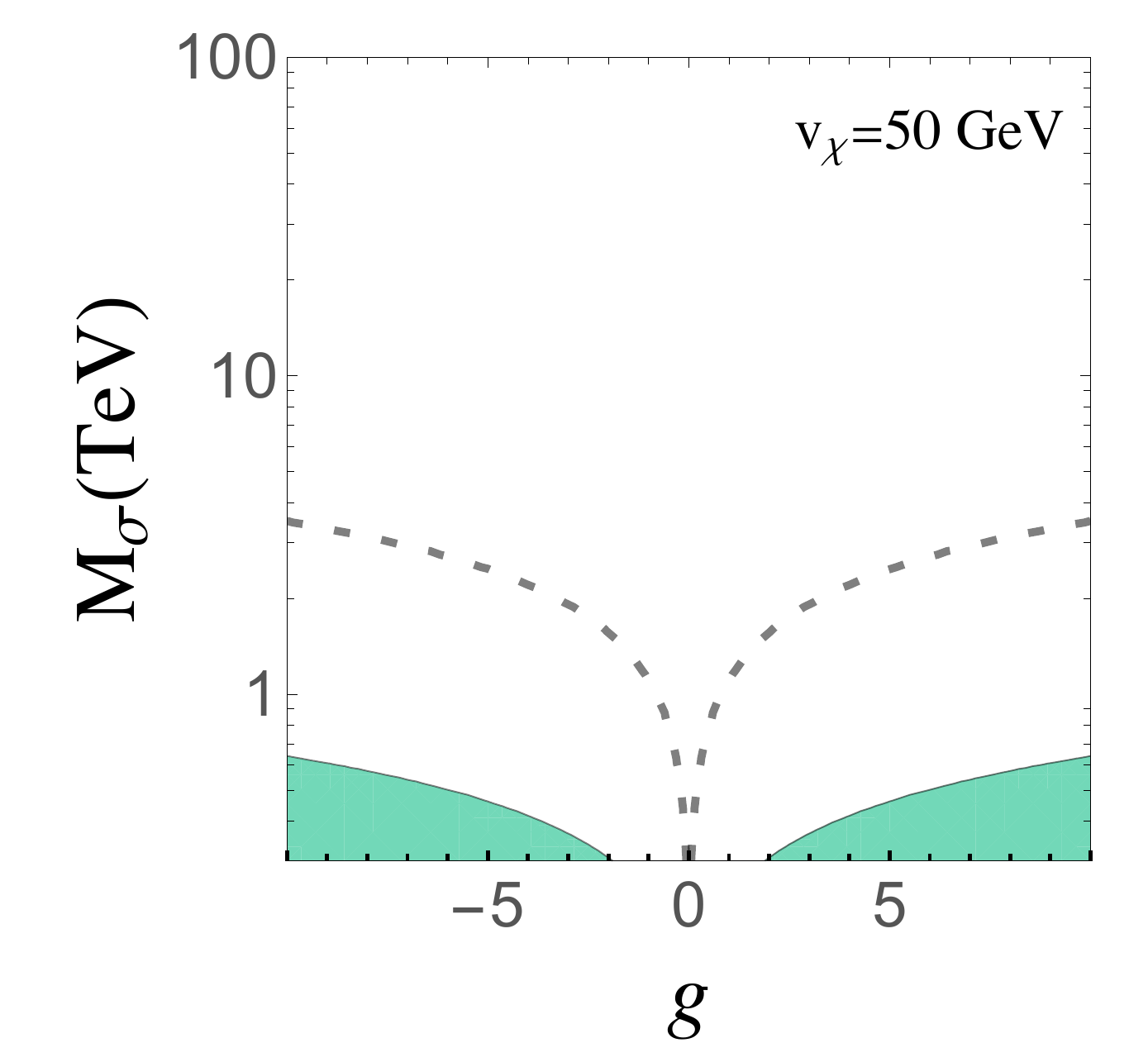} \hspace{-1.95 cm}
\includegraphics[height=5.5cm,keepaspectratio]{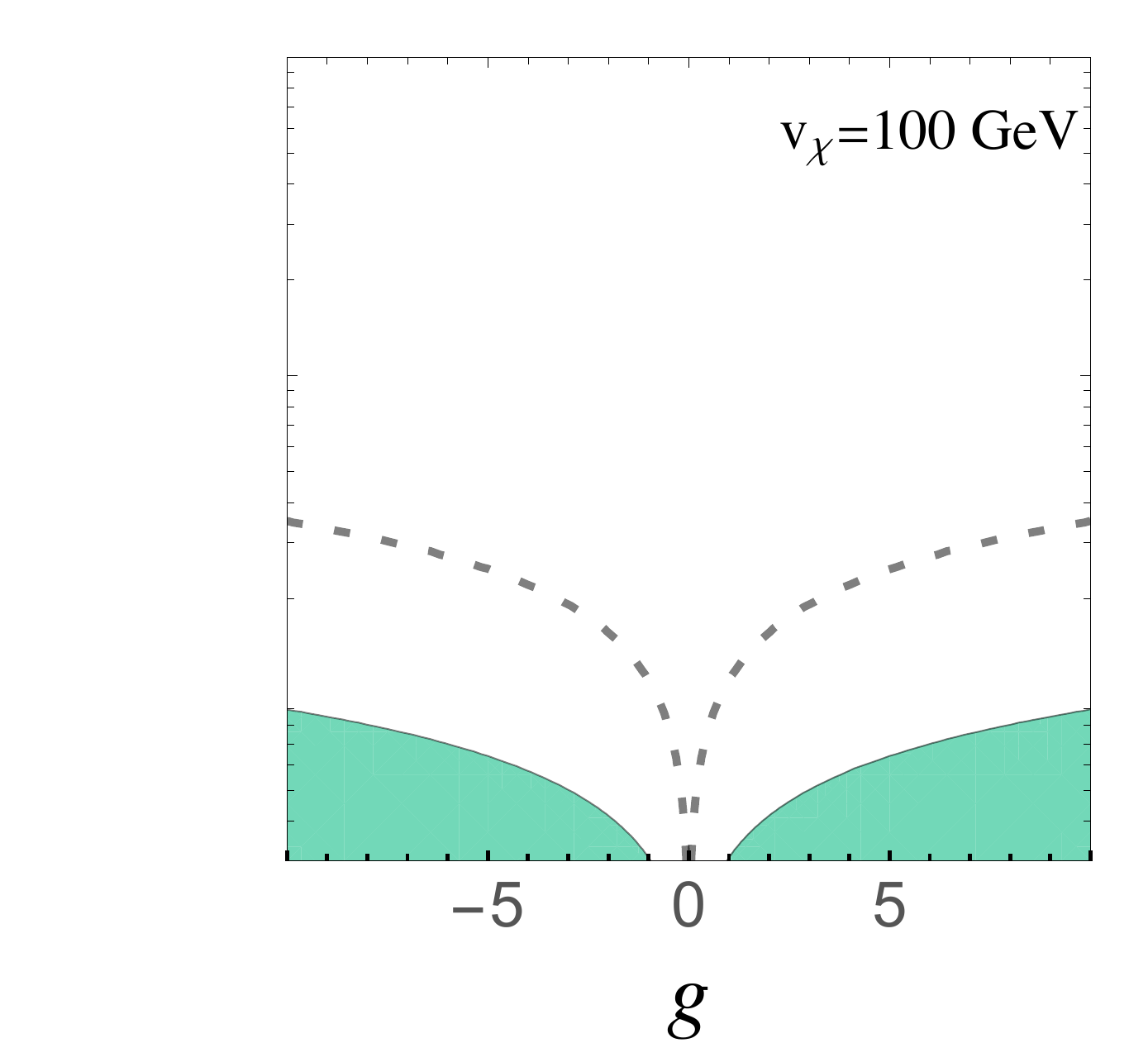} \hspace{-1.95 cm}
\includegraphics[height=5.5cm,keepaspectratio]{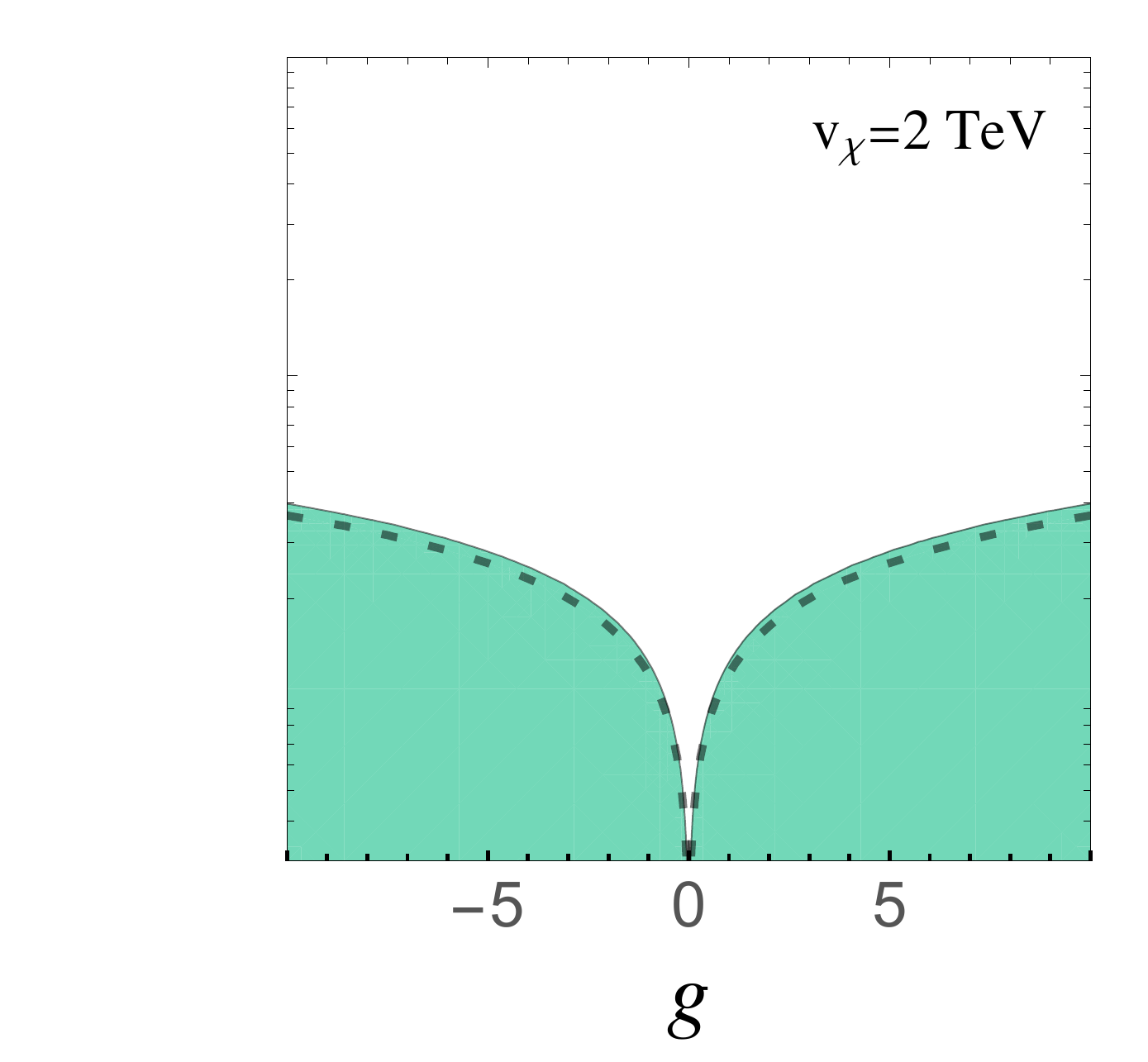}
\caption{\em $M_\sigma$ {\it vs.} $g$ parameter space for three values of $v_\chi$. The green region is excluded by Higgs mixing limits, while the one below the dashed line is ruled out due to the bounds on invisible Higgs decay.}
\label{Fig:SigmaMassPlots}
\end{figure}
\section{MFV embedding}
The Majoron mechanism we have introduced could be included in a larger MFV scheme~\cite{chiv}, based on the flavour symmetry group of the SM~\cite{mfv}, which including RHNs reads $\mathcal{G}_F=U(3)_{q_L}\times U(3)_{u_R}\times U(3)_{d_R}\times U(3)_{\ell_L}\times U(3)_{e_R}\times U(3)_{N_R}$. The Abelian part of this  symmetry group consists of six copies of $U(1)$, two of which could be associated to LN and a Peccei-Quinn symmetry. The former would be broken via a Majoron in the way explained above,  while the latter would also be broken by an axion. 

This setup could possibly offer solutions to several open problems. The Hubble tension would be alleviated by the presence of the Majoron, while strong CP would be solved by the axion in the traditional way.  Besides, the LN and PQ charge assignments could generate, via a Froggatt-Nielsen mechanism, answers to the flavour puzzle, as well as explaining neutrino masses as discussed above. 
\section{Conclusions}
We have explored an SM extension consisting of 3 RHNs and a new scalar, that breaks lepton number symmetry. Two LN choices could provide an improvement of the Hubble tension, via a Majoron, and an explanation for neutrino masses. Both choices satisfy experimental constraints. This model can be included in an MFV framework that possibly solves other open problems of the SM, such as the flavour puzzle or strong CP.

\end{document}